\documentclass[aps,prd,onecolumn,groupedaddress,showpacs,nofootinbib,amssymb]{revtex4}
\usepackage[dvips]{graphicx}
\usepackage{amssymb}
\usepackage{amsmath}
\usepackage{graphicx,color}
\usepackage{amsfonts}
\usepackage{bm}
\usepackage{cancel}
\usepackage{comment}

\newcommand\be{\begin{equation}}
\newcommand\ee{\end{equation}}

\allowdisplaybreaks[4]

\begin{document}

\tolerance=5000

\title{Universal Inflationary Attractors Implications on Static Neutron Stars}
\author{V.K.~Oikonomou,$^{1,2}$\,\thanks{v.k.oikonomou1979@gmail.com}}
 \affiliation{$^{1)}$ Department of Physics, Aristotle University of
Thessaloniki, Thessaloniki 54124,
Greece\\
$^{2)}$ Laboratory for Theoretical Cosmology, Tomsk State
University of Control Systems and Radioelectronics, 634050 Tomsk,
Russia (TUSUR)}

\tolerance=5000

\begin{abstract}
We study static neutron stars in the context of a class of
non-minimally coupled inflationary potentials, the universal
attractors. Universal attractors are known to generate a viable
inflationary era, and they fall into the same category of
inflationary phenomenology as the $R^2$ model and other well-known
cosmological attractors. We present the essential features of
universal attractors in both the Einstein and Jordan frame, and we
extract the Tolman-Oppenheimer-Volkoff equations in the Einstein
frame using the usual notation of theoretical astrophysics. We use
a python 3 based double shooting numerical code for our numerical
analysis and we construct the $M-R$ graphs for the universal
attractor potential, using piecewise polytropic equation of state
the small density part of which is the WFF1 or the APR or the SLy
equation of state. As we show, all the studied cases predict
larger maximum masses for the neutron stars, and all the results
are compatible with the GW170817 constraints imposed on the radii
of the neutron stars.
\end{abstract}

\pacs{04.50.Kd, 95.36.+x, 98.80.-k, 98.80.Cq,11.25.-w}

\maketitle

\section*{Introduction}

The last three decades have brought cosmology and astrophysics to
the mainstream of physics, since the observation of dark energy
\cite{Riess:1998cb} and the direct detection of gravitational
waves \cite{TheLIGOScientific:2017qsa,Abbott:2020khf} have altered
the way of thinking on how the Universe works in small and large
scales. Neutron stars (NS)
\cite{Haensel:2007yy,Friedman:2013xza,Baym:2017whm,Lattimer:2004pg,Olmo:2019flu}
currently are in the interest of many scientific areas, like
nuclear theory
\cite{Lattimer:2012nd,Steiner:2011ft,Horowitz:2005zb,Watanabe:2000rj,Shen:1998gq,Xu:2009vi,Hebeler:2013nza,Mendoza-Temis:2014mja,Ho:2014pta,Kanakis-Pegios:2020kzp},,
high energy physics
\cite{Buschmann:2019pfp,Safdi:2018oeu,Hook:2018iia,Edwards:2020afl,Nurmi:2021xds},
modified gravity
\cite{Astashenok:2020qds,Astashenok:2021peo,Capozziello:2015yza,Astashenok:2014nua,Astashenok:2014pua,Astashenok:2013vza,Arapoglu:2010rz,Panotopoulos:2021sbf,Lobato:2020fxt}
and astrophysics
\cite{Bauswein:2020kor,Vretinaris:2019spn,Bauswein:2020aag,Bauswein:2017vtn,Most:2018hfd,Rezzolla:2017aly,Nathanail:2021tay,Koppel:2019pys}.

There is strong evidence coming from the observations on dark
energy that modified gravity in its various forms
\cite{Nojiri:2017ncd,Nojiri:2009kx,Capozziello:2011et,Capozziello:2010zz,Nojiri:2006ri,
Nojiri:2010wj,delaCruzDombriz:2012xy,Olmo:2011uz} actually plays a
fundamental role on large scales. Also at the astrophysical level,
it is possible to generate large or extremely large neutron star
masses and solve several fundamental equation of state (EoS)
related problems of neutron stars
\cite{Astashenok:2014nua,Astashenok:2014pua}. Hence it is probable
that general relativity (GR) by itself may not suffice to describe
NSs, hence some extension of GR might be compelling. In this work
we shall consider NSs in hydrodynamic equilibrium in the context
of non-minimally coupled scalar-tensor theories. This subject is
very well studied in the theoretical astrophysics literature, see
Refs.
\cite{Pani:2014jra,Staykov:2014mwa,Horbatsch:2015bua,Silva:2014fca,Doneva:2013qva,Xu:2020vbs,Salgado:1998sg,Shibata:2013pra,Arapoglu:2019mun,Ramazanoglu:2016kul,AltahaMotahar:2019ekm,Chew:2019lsa,Blazquez-Salcedo:2020ibb,Motahar:2017blm}
for an important stream of articles on this subject. We shall
consider some not so well known in the theoretical astrophysics
literature non-minimal coupled theories, those of cosmological
attractors
\cite{alpha1,alpha2,alpha3,alpha4,alpha5,alpha6,alpha7,alpha7a,alpha8,alpha9,alpha10,alpha11,alpha12,alpha13,alpha14,alpha15,alpha16,alpha17,alpha18,alpha19,alpha20,alpha21,alpha22,alpha23,alpha24,alpha25,alpha26,alpha27,alpha28,alpha29,alpha30,alpha31,alpha32,alpha33,alpha34,alpha35,alpha36,alpha37}.
Specifically, in this work we shall consider the class known as
universal attractors \cite{alpha7a}, and we shall investigate the
implications of such non-minimally coupled scalar field theories
on static NSs in the Einstein frame. These models are known to
provide a uniform inflationary phenomenology and belong to the
larger class of cosmological attractors, which provide a viable
inflationary phenomenology compatible with the latest Planck data
\cite{Akrami:2018odb}. We shall solve numerically the
Tolman-Oppenheimer-Volkoff (TOV) equations, using an ``LSODA''
integrator python 3 based code, which is a modification of
\cite{niksterg}, and with regard to the EoS for the nuclear
matter, we shall assume that the EoS is a piecewise polytropic EoS
\cite{Read:2008iy,Read:2009yp}, with the low density part being
the WFF1 \cite{Wiringa:1988tp}, the SLy \cite{Douchin:2001sv}, of
the APR EoS \cite{Akmal:1998cf}. With regard to the mass of the
NS, we shall find the numerical value of the Einstein frame
Arnowitt-Deser-Misner (ADM) mass \cite{Arnowitt:1960zzc} and from
it we shall calculate the Jordan frame mass numerically.

This paper is organized as follows: In section I we review the
essential features of the universal attractors in the context of
cosmology, and we shall demonstrate how these models provide a
viable inflationary era. In section II we discuss these models in
the context of theoretical astrophysics notation and physical
units, and we solve numerically the TOV equations for the three
distinct EoSs. In the same section we qualitatively discuss the
phenomenological features of the NSs for the universal attractor
potentials. Finally the conclusions follow in the end of the
paper.

\section{Essential Features of Universal Attractor Theories}

Universal attractors belong to a large class of cosmological
attractors studied in Refs.
\cite{alpha1,alpha2,alpha3,alpha4,alpha5,alpha6,alpha7,alpha7a,alpha8,alpha9,alpha10,alpha11,alpha12,alpha13,alpha14,alpha15,alpha16,alpha17,alpha18,alpha19,alpha20,alpha21,alpha22,alpha23,alpha24,alpha25,alpha26,alpha27,alpha28,alpha29,alpha30,alpha31,alpha32,alpha33,alpha34,alpha35,alpha36,alpha37}.
All these cosmological attractor models originate from various
forms of Jordan frame non-minimally coupled scalar theories, but
the Einstein frame counterpart yield a quite similar inflationary
phenomenology, for generic non-minimal coupling. In this section
we shall briefly demonstrate how the universal attractors
inflationary theory is obtained. The notation we shall use is
frequently used in cosmological contexts, so we use natural units
for this section. In the next section where we study the Einstein
frame NS phenomenology, we switch to Geometrized units. The
conventions and formalism of conformal transformations in
cosmological contexts, we refer the reader to Refs.
\cite{Kaiser:1994vs,valerio}.

We start of with the Jordan frame action of a non-minimally
coupled scalar field $\phi$,
\begin{equation}\label{c1}
\mathcal{S}_J=\int
d^4x\Big{[}f(\phi)R-\frac{\omega(\phi)}{2}g^{\mu
\nu}\partial_{\mu}\phi\partial_{\nu}\phi-U(\phi)\Big{]}+S_m(g_{\mu
\nu},\psi_m)\, ,
\end{equation}
with $\psi_m$ denoting the Jordan frame perfect matter fluids,
with $P$ and energy density $\epsilon$. The universal attractors
in the Jordan frame have the following non-minimal coupling,
\begin{equation}\label{fjordanuni}
f(\phi)=\frac{M_p^2}{2}\left(1-\xi \phi^2 \right)\, ,
\end{equation}
where $\xi$ is a positive constant coupling, and the reduced
Planck mass is defined as follows,
\begin{equation}\label{c3}
M_p=\frac{1}{\sqrt{8\pi G}}\, ,
\end{equation}
where $G$ is the gravitational constant of Newtonian gravity.
Moreover, the universal attractors in the Jordan frame have the
following scalar potential,
\begin{equation}\label{unipotentialjordan}
U(\phi)=U_0f^2(\phi)\left(\frac{\phi}{M_p}\right)^{2n}\, ,
\end{equation}
with $n$ being some positive number. Now if the following
conformal transformation is performed in the Jordan frame action
with metric $g_{\mu \nu}$,
\begin{equation}\label{c4}
\tilde{g}_{\mu \nu}=\Omega^2g_{\mu \nu}\, ,
\end{equation}
we obtain the Einstein frame action with metric $\tilde{g}_{\mu
\nu}$, where the ``tilde'' will denote the Einstein frame
quantities. If we use \cite{Kaiser:1994vs,valerio},
\begin{equation}\label{c6}
\Omega^2=\frac{2}{M_p^2}f(\phi)\, ,
\end{equation}
we may obtain a minimal coupled scalar theory in the Einstein
frame,
\begin{equation}\label{c12}
\mathcal{S}_E=\int
d^4x\sqrt{-\tilde{g}}\Big{[}\frac{M_p^2}{2}\tilde{R}-\frac{\zeta
(\phi)}{2} \tilde{g}^{\mu \nu }\tilde{\partial}_{\mu}\phi
\tilde{\partial}_{\nu}\phi-V(\phi)\Big{]}+S_m(\Omega^{-2}\tilde{g}_{\mu
\nu},\psi_m)\, ,
\end{equation}
where,
\begin{equation}\label{c14}
\zeta
(\phi)=\frac{M_p^2}{2}\Big{(}\frac{3\Big{(}\frac{df}{d\phi}\Big{)}^2}{f^2}+\frac{2\omega(\phi)}{f}\Big{)}\,
,
\end{equation}
and the potential $V(\phi)$ is written in terms of the Jordan
frame potential as,
\begin{equation}\label{c13}
V(\phi)=\frac{U(\phi)}{\Omega^4}\, ,
\end{equation}
thus in view of Eqs. (\ref{unipotentialjordan}) and (\ref{c13}),
the Einstein frame potential for the universal attractors reads,
\begin{equation}\label{unipotentialjordan}
V(\phi)=\frac{U_0 M_p^4}{4}\left(\frac{\phi}{M_p}\right)^{2n}\, .
\end{equation}
The Einstein frame scalar field $\phi$ can be made canonical by
using the following transformation,
\begin{equation}\label{c15}
\Big{(}\frac{d\varphi}{d \phi}\Big{)} =\sqrt{\zeta(\phi)}\, ,
\end{equation}
hence the Einstein frame action becomes,
\begin{equation}\label{c17}
\mathcal{S}_E=\int
d^4x\sqrt{-\tilde{g}}\Big{[}\frac{M_p^2}{2}\tilde{R}-\frac{1}{2}\tilde{g}^{\mu
\nu } \tilde{\partial}_{\mu}\varphi
\tilde{\partial}_{\nu}\varphi-V(\varphi)\Big{]}+S_m(\Omega^2\tilde{g}_{\mu
\nu},\psi_m)\, .
\end{equation}
The Einstein frame matter fluids are coupled to the conformal
factor so these are not perfect, because the energy momentum
tensor satisfies,
\begin{equation}\label{c24}
\tilde{\partial}^{\mu}\tilde{T}_{\mu \nu}=-\frac{d}{d\varphi}[\ln
\Omega]\tilde{T}\tilde{\partial}_{\nu}\phi\, .
\end{equation}
Hereafter we shall assume that,
\begin{equation}\label{omegacons1}
\Omega(\phi)\ll \frac{3 M_p^2}{2}\Omega'(\phi)\, ,
\end{equation}
which by using the analytic form of $\Omega$ for the universal
attractors, the above condition can be written as follows,
\begin{equation}\label{conditionomega2}
1-\frac{\xi \phi^2}{M_p^2}\ll \frac{6 \xi^2 \phi^2}{M_p^2}\, .
\end{equation}
By substituting the analytic form of $f(\phi)$ from Eq.
(\ref{fjordanuni}) into Eq. (\ref{c14}), we have,
\begin{equation}\label{c14proeq}
\frac{d\varphi}{d\phi}=\frac{\sqrt{1+\frac{\xi
\phi^2}{M_p^2}+\frac{6\xi^2\phi^2}{M_p^2}}}{1+\frac{\xi
\phi^2}{M_p^2}}\, ,
\end{equation}
so in view of the assumption (\ref{omegacons1}), we easily obtain
from Eq. (\ref{c14proeq}) the following,
\begin{equation}\label{finalphioff}
\varphi=-\sqrt{\frac{3}{2}}M_p\ln \left(1-\frac{\xi \phi^2}{M_p^2}
\right)\, ,
\end{equation}
or equivalently,
\begin{equation}\label{finalfofvarphi}
\frac{\phi^2}{M_p^2}=\frac{1-e^{-\sqrt{\frac{2}{3}}\frac{\varphi}{M_p}}}{\xi}\,
.
\end{equation}
Hence, by substituting Eq. (\ref{finalfofvarphi}) in Eq.
(\ref{unipotentialjordan}) we finally obtain the Einstein frame
scalar potential in terms of the canonical scalar field $\varphi$,
\begin{equation}\label{finalpotentialeinsteininflation}
V(\varphi)=\frac{U_0
M_p^4}{4\xi^n}\left(1-e^{-\sqrt{\frac{2}{3}}\frac{\varphi}{M_p}}\right)^{n}\,
.
\end{equation}
Let us set for convenience $V_0=\frac{U_0 M_p^4}{4\xi^n}$, and by
taking into the Planck constraints \cite{Akrami:2018odb} on the
amplitude $\Delta_s^2$ of single canonical scalar field
fluctuations,
\begin{equation}\label{scalarampconst}
\Delta_s^2=2.2\times 10^{-9}\, ,
\end{equation}
where $\Delta_s^2$ is,
\begin{equation}\label{scalaramp}
\Delta_s^2=\frac{1}{24\pi^2}\frac{V(\varphi_f)}{M_p^4}\frac{1}{\epsilon(\varphi_f)}\,
,
\end{equation}
the parameter $V_0$ is constrained to be,
\begin{equation}\label{tilde}
V_0\sim 9.6\times 10^{-11}\, M_p^4\, .
\end{equation}
Let us elaborate further on the constraint we just quoted, namely
the parameter $V_0$. This parameter is constrained by the Planck
data in a model independent way, using the BK15 constraint on $r$
(which is $r=16 \epsilon$), see for example Eq. (32) of the Planck
2018 constraints on inflation, page 14 \cite{Akrami:2018odb}. In
that equation the tensor-to-scalar ratio is used, while we
replaced $r=16\epsilon$ in Eq. (\ref{scalaramp}). Also note that
in our notation the amplitude of the scalar fluctuations is
$\Delta_s^2$ while in the Planck data this is denoted as $A_s$.
Therefore the constraint of Eq. (32) of Ref. \cite{Akrami:2018odb}
is equivalent to our constraint (upper bound) Eq.
(\ref{scalarampconst}), which if we substitute the maxim allowed
values of the slow-roll index $\epsilon (\varphi_f)$ (or
equivalently the maximum allowed value of the tensor-to-scalar
ratio), we obtain the constraint (\ref{tilde}) of our paper. This
is obtained in a general and model-independent way and does not
rely on the specifics of the model used, it is solely based on the
Planck constraints on canonical scalar field inflation.

Clearly the constraint on $V_0$ is an upper bound and we thus
focused on this upper bound case in our paper. Definitely the
parameter $V_0$ can take smaller values, but we used the upper
bounds values for the slow-roll parameters $\epsilon$ and $r$,
thus we focused our analysis on the maximum value for the scale of
inflation $V_0$. In principle one could use lower values for the
scale of inflation, until for example
$\frac{V_0^{\frac{1}{2}}\kappa}{3}\sim 10^{12}\,$GeV, which is the
low-scale inflation constraint (we used the slow-roll relation
$\frac{3H^2}{\kappa^2}\sim V$), but this would not change
drastically the parameter $V_0$, plus one should explain how the
low-scale inflation scenario occurs. Hence in our approach we used
the most plausible values for the scale of inflation, inherently
connected to the ordinary scale of inflation, not the low-scale of
inflation.

Note that $\varphi_f$ and $\epsilon$ in Eq. (\ref{scalaramp}) are
the value of the canonical scalar field in the Einstein frame at
the end of inflation and the first slow-roll index. The canonical
scalar theory in the Einstein frame with the potential
(\ref{finalpotentialeinsteininflation}), which has a resulting
action in the Einstein frame,
\begin{equation}\label{einsteinframeaction}
\mathcal{S}_E=\int
d^4x\sqrt{-\tilde{g}}\Big{[}\frac{M_p^2}{2}\tilde{R}-\frac{1}{2}\tilde{g}^{\mu
\nu } \tilde{\partial}_{\mu}\varphi
\tilde{\partial}_{\nu}\varphi-V_0\left(1-e^{-\sqrt{\frac{2}{3}}\frac{\varphi}{M_p}}\right)^{n}\Big{]}\,
,
\end{equation}
yields a viable inflationary phenomenology and has an attractor
behavior resulting to the following spectral index of primordial
scalar curvature perturbations $n_s$ and tensor-to-scalar ratio,
at leading order in the large $e$-foldings number $N$,
\begin{equation}\label{spectralindexsmallalpha}
n_s=1-\frac{2}{N}\, ,\,\,\,r=\frac{12}{N^2}\, .
\end{equation}
The above observational indices for inflation are identical to the
ones corresponding to the $R^2$ model and other inflationary
phenomenological models. A useful expression for the action
(\ref{einsteinframeaction}) is the following,
\begin{equation}\label{einsteinframeactioninflationns}
\mathcal{S}_E=\int d^4x\sqrt{-\tilde{g}}\Big{[}\frac{1}{16\pi
G}\tilde{R}-\frac{1}{2}\tilde{g}^{\mu \nu }
\tilde{\partial}_{\mu}\varphi
\tilde{\partial}_{\nu}\varphi-\frac{16\pi G V(\varphi)}{16\pi
G}\Big{]}\, ,
\end{equation}
and recall that $M_p^2=\frac{1}{8\pi G}$. The above form of the
action is convenient for the universal attractor theory in the
Einstein frame in the context of theoretical astrophysics
notation. Also let us comment that the action
(\ref{einsteinframeaction}) us identical to the one corresponding
to the $R^2$ model for $n=2$, and this is the case we shall also
study. However, the two theories yield only identical inflation
but the two theories are not the same because the conformal factor
$\Omega$ and the resulting coupling to the matter fluids are not
the same, so these two theories look like the same but are not the
same. We evince this feature in the next section.

\section{Neutron Stars in the Einstein Frame with Universal Attractors}

Let us now study the universal attractor potentials in
astrophysical contexts. We shall use Geometrized units $G=c=1$,
and also we adopt the notation and conventions of Ref.
\cite{Pani:2014jra}. The Jordan frame action of a non-minimally
coupled scalar field in the presence of perfect matter fluids
$\psi_m$ is,
\begin{equation}\label{ta}
\mathcal{S}=\int
d^4x\frac{\sqrt{-g}}{16\pi}\Big{[}f(\phi)R-\frac{1}{2}g^{\mu
\nu}\partial_{\mu}\phi\partial_{\nu}\phi-U(\phi)\Big{]}+S_m(\psi_m,g_{\mu
\nu})\, .
\end{equation}
By conformally transforming the above action using,
\begin{equation}\label{ta1higgs}
\tilde{g}_{\mu \nu}=A^{-2}g_{\mu \nu}\,
,\,\,\,A(\phi)=f^{-1/2}(\phi)\, ,
\end{equation}
we obtain the Einstein frame action which is,
\begin{equation}\label{ta5higgs}
\mathcal{S}=\int
d^4x\sqrt{-\tilde{g}}\Big{(}\frac{\tilde{R}}{16\pi}-\frac{1}{2}
\tilde{g}_{\mu \nu}\partial^{\mu}\varphi
\partial^{\nu}\varphi-\frac{V(\varphi)}{16\pi}\Big{)}+S_m(\psi_m,A^2(\varphi)g_{\mu
\nu})\, ,
\end{equation}
where $\varphi$ is the Einstein frame canonical scalar field,
which is related to the scalar field $\phi$ as follows,
\begin{equation}\label{ta4higgs}
\frac{d \varphi }{d \phi}=\frac{1}{\sqrt{4\pi}}
\sqrt{\Big{(}\frac{3}{4}\frac{1}{f^2}\Big{(}\frac{d
f}{d\phi}\Big{)}^2+\frac{1}{4f}\Big{)}}\, ,
\end{equation}
while the potential $\mathcal{V}(\varphi)$ is,
\begin{equation}\label{potentialns1}
V(\varphi)=\frac{U(\phi)}{f^2}\, .
\end{equation}
Now the universal attractors case corresponds to the choices,
\begin{equation}\label{funins}
f(\phi)=1-\xi \phi^2\, ,\,\,\,U(\phi)=\mathcal{U}_0f^2(\phi)
\phi^{2n}\, ,
\end{equation}
and for these choices, Eq. (\ref{ta4higgs}) becomes,
\begin{equation}\label{neweqn}
\frac{d \varphi}{d
\phi}=\frac{1}{\sqrt{16\pi}}\frac{\sqrt{1-\xi\phi^2+12\xi^2\phi^2}}{1-\xi\phi^2}\,
.
\end{equation}
Now, for the case and notation at hand, the assumption of Eq.
(\ref{omegacons1}) reads,
\begin{equation}\label{mainassumptionuniversal}
1-\xi\phi^2\ll 12\xi^2\phi^2\, ,
\end{equation}
and in view of this assumption, we can integrate Eq.
(\ref{ta4higgs}) to obtain,
\begin{equation}\label{fophiequn}
\phi^2=\frac{1}{\xi}\left(1-e^{-4\sqrt{\frac{\pi}{3}}}\right)\, ,
\end{equation}
and now due to Eq. (\ref{ta1higgs}) the conformal factor
$A(\varphi)$ reads,
\begin{equation}\label{conformalfactor}
A(\varphi)=e^{2\sqrt{\frac{\pi}{3}}\varphi}\, .
\end{equation}
In addition, a useful quantity which shall be used in the
following is the function $\alpha (\varphi )$ defined as follows,
\begin{equation}\label{alphaofvarphigeneraldef}
\alpha(\varphi)=\frac{d A(\varphi)}{d \varphi}\, ,
\end{equation}
hence in the case of universal attractors,
\begin{equation}\label{alphaofphifinalintermsofvarphi}
a(\varphi)=\alpha=2\sqrt{\frac{\pi}{3}}\, .
\end{equation}
It is worth noting that when $f(\phi)\to \infty$ then $A(\phi)\to
0$, thus at this limit the Einstein and Jordan frame are not
equivalent, see for example \cite{Bhattacharya:2020wdl}, however
we do not have to worry for this limit, since during inflation,
the values of the scalar field are of the order of the Planck
mass, while even at astrophysical contexts, such as interior and
exterior of scalar-tensor neutron stars, the values of the scalar
field are significantly smaller than the Planck mass. Also, using
(\ref{fophiequn}) and (\ref{alphaofphifinalintermsofvarphi}) the
potential in the Einstein frame takes the final form,
\begin{equation}\label{finaleinsteinframepotentialvaprhi}
V=\mathcal{V}_0\left( 1-e^{-2\alpha \varphi }\right)^n\, ,
\end{equation}
where $\mathcal{V}_0=\frac{\mathcal{U}_0}{\xi^n}$. From this point
we shall assume that $n=2$, thus let us use the constraints on
$V_0$ from the previous section to determine the values of
$\mathcal{V}_0$ always working in Geometrized units with $G=1$. By
comparing the actions (\ref{einsteinframeactioninflationns}) and
(\ref{finaleinsteinframepotentialvaprhi}), we have
$\mathcal{V}_0=16\pi V_0$, so $\mathcal{V}_0\simeq 7.62094\times
10^{-12}$. Hence, since
$\mathcal{V}_0=\frac{\mathcal{U}_0}{\xi^2}$, we can choose
$\mathcal{U}_0=1$ and $\xi=36.2239\times 10^{4}$, and we can
observe that for this choice we shall also explicitly check
whether the constraint (\ref{mainassumptionuniversal}) holds true,
in the Jordan frame.
\begin{figure}[h!]
\centering
\includegraphics[width=23pc]{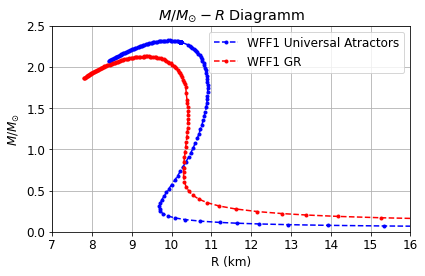}
\includegraphics[width=23pc]{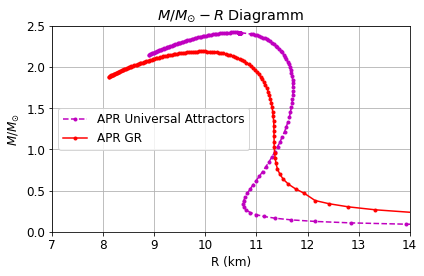}
\includegraphics[width=23pc]{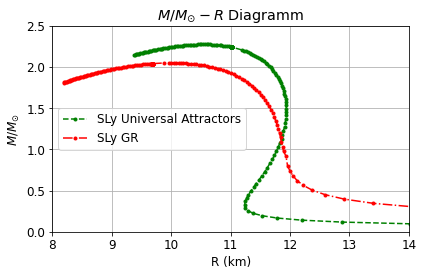}
\caption{$M-R$ graphs for the universal attractor model compared
to the GR case, for the WFF1 EoS (upper), the APR EoS (middle) and
the SLy EoS (bottom). The GW170817 indicate that for a NS of mass
$M\sim 1.6 M_{\odot}$, the predicted radii for the universal
attractor models must be larger than $R=10.68^{+15}_{-0.04}$km.
Also the GW170817 event indicates that the radii corresponding to
the maximum mass must be larger than $R=9.6^{0.14}_{-0.03}$km. All
the GW170817 constraints are satisfied as it can be seen in the
plots. } \label{plot1}
\end{figure}
We shall consider static NSs in the Einstein frame. Thus, the
spherically symmetric static metric which describes the NS used in
this paper is, \begin{equation}\label{tov1}
ds^2=-e^{\nu(r)}dt^2+\frac{dr^2}{1-\frac{2
m(r)}{r}}+r^2(d\theta^2+\sin^2\theta d\phi^2)\, ,
\end{equation}
where the function $m(r)$ stands for the gravitational mass of the
NS with circumferential radius $r$. It is worth discussing here
the issue of exterior and interior spacetime for the static
neutron star. In standard GR contexts where the scalar field is
absent, the exterior spacetime of the neutron star is
Schwarzschild, however in the presence of the scalar field the
spacetime is uniformly given by the metric (\ref{tov1}), see for
example \cite{Pani:2014jra}. It is the aim of any study in
scalar-tensor astrophysics to find, numerically, the metric
functions $\nu(r)$ and $\frac{1}{1-\frac{2 m(r)}{r}}$. Obviously
no matching conditions are required at the surface of the star,
because the TOV equations will yield a continuous solution for the
metric functions, starting from the interior of the star,
extending to the surface of the star and these solutions will
describe the star asymptotically, thus at $r\to \infty$. At
exactly this point, the numerical infinity, the exterior spacetime
will be a Schwarzschild spacetime, and this is the only condition
imposed on the metric functions $\nu(r)$ and $\frac{1}{1-\frac{2
m(r)}{r}}$. These have to be matched with the corresponding
Schwarzschild ones. The solutions for the metric functions
$\nu(r)$ and $\frac{1}{1-\frac{2 m(r)}{r}}$ will be obtained
numerically by solving the TOV equations continuously from the
interior until the numerical infinity. The difference between the
interior of the star and the exterior is that the exterior does
not have contribution from the matter of the star, thus the
pressure and the energy density outside the star are zero.
However, the potential is not switched off, thus it affects the
metric function solutions even outside the star. This is the major
difference of scalar-tensor gravity and GR for astrophysical
compact objects. For the numerical procedure, it is important to
match the metric function solutions asymptotically with the
Schwarzschild ones, and thus a double shooting method is required
for this, in order to find the optimal initial conditions at the
center of the star which guarantee that the spacetime
asymptotically will be matched with the Schwarzschild. The double
shooting method we used guarantees that the spacetime at numerical
infinity is Schwarzschild. Now we can derive the TOV equations,
which are \cite{Pani:2014jra},
\begin{equation}\label{tov2}
\frac{d m}{dr}=4\pi r^2
A^4(\varphi)\varepsilon+\frac{r}{2}(r-2m(r))\omega^2+4\pi
r^2V(\varphi)\, ,
\end{equation}
\begin{equation}\label{tov3}
\frac{d\nu}{dr}=r\omega^2+\frac{2}{r(r-2m(r))}\Big{[}4\pi
A^4(\varphi)r^3P-4\pi V(\varphi)
r^3\Big{]}+\frac{2m(r)}{r(r-2m(r))}\, ,
\end{equation}
\begin{equation}\label{tov4}
\frac{d\omega}{dr}=\frac{4\pi r
A^4(\varphi)}{r-2m(r)}\Big{(}\alpha(\varphi)(\epsilon-3P)+
r\omega(\epsilon-P)\Big{)}-\frac{2\omega
(r-m(r))}{r(r-2m(r))}+\frac{8\pi \omega r^2 V(\varphi)+r\frac{d
V(\varphi)}{d \varphi}}{r-2 m(r)}\, ,
\end{equation}
\begin{equation}\label{tov5}
\frac{dP}{dr}=-(\epsilon+P)\Big{[}\frac{1}{2}\frac{d\nu}{dr}+\alpha
(\varphi)\omega\Big{]}\, ,
\end{equation}
\begin{equation}\label{tov5newfinal}
\omega=\frac{d \varphi}{dr}\, ,
\end{equation}
with the function $\alpha (\varphi)$ is defined in Eq.
(\ref{alphaofphifinalintermsofvarphi}). Note that the pressure and
the energy density $P$ and $\epsilon$, are Jordan frame
quantities. Also the interior and the interior spacetime are
uniformly described by the metric (\ref{tov1}), no discrimination
to exterior and interior spacetimes is done here, like in GR. The
TOV equations for the spacetime outside the star can be derived by
setting $P=0$ and $\epsilon=0$, which shows the absence of matter
outside the star. However, the potential is still present thus it
affects the star beyond its surface. This is exactly why no
matching outside the star is needed at the surface. The numerical
solutions of the TOV equations will yield continuous uniform
solutions for the metric functions, and the only condition
required is that asymptotically, these solutions must become
identical to the corresponding Schwarzschild ones.

The initial conditions for the TOV equations are,
\begin{equation}\label{tov8}
P(0)=P_c\, ,\,\,\,m(0)=0\, , \,\,\,\nu(0)\, ,=-\nu_c\, ,
\,\,\,\varphi(0)=\varphi_c\, ,\,\,\, \omega (0)=0\, .
\end{equation}
The condition $m(0)=0$ means that the gravitational mass for zero
radius is zero. This however does not make the metric function
$\nu(r)$ to be non-zero at zero radius. The exact value
$\nu(0)=-\nu_c$ will be obtained by the double shooting method.
With regard to the EoS, we shall use a piecewise polytropic EoS
\cite{Read:2008iy,Read:2009yp} (see also \cite{niksterg}), with
the low density parameters corresponding to the SLy, WFF1 or the
APR EoSs. For the piecewise polytropic EoS, the relation between
the energy density and pressure is,
\begin{equation}\label{pp4}
\epsilon(\rho) = (1+\alpha_i)\rho +
\frac{K_i}{\Gamma_i-1}\rho^{\Gamma_i}\, ,\,\,\,\alpha_i =
\frac{\epsilon(\rho_{i-1})}{\rho_{i-1}} -1 -
\frac{K_i}{\Gamma_i-1}\rho_{i-1}^{\Gamma_i-1}\, ,
\end{equation}
where the $i$ refers to the three different pieces of the
polytropic equation of state. Let us elaborate on this further,
the energy density and the pressure in each of the three piecewise
density intervals $\rho_{i-1} \leq \rho \leq \rho_i$ satisfy the
polytropic relation,
\begin{equation}\label{pp1}
P = K_i\rho^{\Gamma_i}\, ,
\end{equation}
and we only have to require that continuity is needed at the
crossing points of each of the three pieces. Particularly, at the
crossing points we must have,
\begin{equation}\label{pp2}
P(\rho_i) = K_i\rho^{\Gamma_i} = K_{i+1}\rho^{\Gamma_{i+1}}\, ,
\end{equation}
and from the above relations, the parameters $K_2$ and $K_3$ are
obtained, given $K_1, \Gamma_1, \Gamma_2, \Gamma_3$, or
equivalently, given the initial pressure $p_1$ and for given
parameters $\Gamma_2$, and $\Gamma_3$, which are not chosen
arbitrarily.  Upon integrating the first thermodynamic law for
barotropic fluids,
\begin{equation}\label{pp3}
d\frac{\epsilon}{\rho} = - P d\frac{1}{\rho}\, ,
\end{equation}
in conjunction with the continuity requirement at each piece of
the polytropic, yields Eq. (\ref{pp4}). Now let us discuss the
gravitational mass issue for the NS. The gravitational mass of the
NS which we shall consider is the ADM mass in the Jordan frame.
Thus when we extract the numerical results, we need to transform
the obtained Einstein frame mass to the Jordan frame. We define
the auxiliary functions $\mathcal{K}_E$ and $\mathcal{K}_J$ in
Geometrized units,
\begin{equation}\label{hE}
\mathcal{K}_E=1-\frac{2 m}{r_E}\, ,\,\,\,\mathcal{K}_J=1-\frac{2
m_J}{r_J}\, ,
\end{equation}
which are basically the metric functions in the Einstein and
Jordan frames, with $m$ and $m_J$ being the gravitational mass
confined in a radius $r$. The metric functions along with the
metric radius parameter in the two frames are related as follows,
\begin{equation}\label{hehjrelation}
\mathcal{K}_J=A^{-2}\mathcal{K}_E\, ,\,\,\,r_J=A r_E\, ,
\end{equation}
and in addition, the Jordan and Einstein frame ADM masses are,
\begin{equation}\label{jordaframemass1}
M_J=\lim_{r\to \infty}\frac{r_J}{2}\left(1-\mathcal{K}_J \right)
\, ,\,\,\,M_E=\lim_{r\to \infty}\frac{r_E}{2}\left(1-\mathcal{K}_E
\right) \, .
\end{equation}
Taking the asymptotic limit of Eq. (\ref{hehjrelation}), we
obtain,
\begin{equation}\label{asymptotich}
\mathcal{K}_J(r_E)=\left(1+\alpha(\varphi(r_E))\frac{d \varphi}{d
r}r_E \right)^2\mathcal{K}_E(\varphi(r_E))\, ,
\end{equation}
where $r_E$ denotes the radius in the Einstein frame
asymptotically (not at the numerical infinity though, slightly
smaller) and in addition $\frac{d\varphi }{dr}=\frac{d\varphi
}{dr}\Big{|}_{r=r_E}$. Upon combining Eqs.
(\ref{hE})-(\ref{asymptotich}) we acquire \cite{Odintsov:2021qbq},
\begin{equation}\label{jordanframeADMmassfinal}
M_J=A(\varphi(r_E))\left(M_E-\frac{r_E^{2}}{2}\alpha
(\varphi(r_E))\frac{d\varphi
}{dr}\left(2+\alpha(\varphi(r_E))r_E\frac{d \varphi}{dr}
\right)\left(1-\frac{2 M_E}{r_E} \right) \right)\, .
\end{equation}
Finally, the Jordan and Einstein frame radii of the NS are related
as follows,
\begin{equation}\label{radiussurface}
R=A(\varphi(R_s))\, R_s\, .
\end{equation}
Hereafter we shall identify with $M$ the Jordan frame mass of Eq.
(\ref{jordanframeADMmassfinal}), that is $M=M_J$ measured in solar
masses, and the radius in the Jordan frame $R$ shall be expressed
in kilometers.
\begin{figure}[h!]
\centering
\includegraphics[width=23pc]{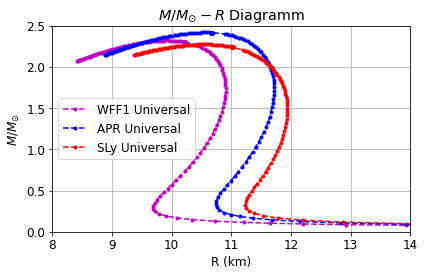}
\caption{Comparison of the $M-R$ graphs for the universal
attractor model for the WFF1, APR and SLy EoSs. The GW170817
indicate that for a NS of mass $M\sim 1.6 M_{\odot}$, the
predicted radii for the universal attractor models must be larger
than $R=10.68^{+15}_{-0.04}$km. Also the GW170817 event indicates
that the radii corresponding to the maximum mass must be larger
than $R=9.6^{0.14}_{-0.03}$km. All the GW170817 constraints are
satisfied as it can be seen in the plots.} \label{plot2}
\end{figure}
At this point we shall proceed to the presentation of our
numerical analysis of the TOV equations. We shall solve the TOV
equations numerically using a python 3 based numerical code which
is a variant form of the pyTOV-STT code \cite{niksterg} using the
``LSODA'' integrator, in order to extract the Jordan frame mass
and the circumferential radius of the NS. The method includes a
double shooting method in order to find the optimal values for the
the metric function $\nu_c$ and for the scalar field $\varphi_c$
at the center of the NS, which make the values of the scalar field
and of the metric function to vanish at the numerical infinity,
with the latter being chosen in kilometers to be $r\sim
67.94378528694695$ km.
\begin{table}[h!]
  \begin{center}
    \caption{\emph{\textbf{Masses with $\frac{M}{M_{\odot}}\simeq 1$ and the Corresponding Radii of Static NS for the Universal Attractors  and for GR}}}
    \label{table2}
    \begin{tabular}{|r|r|r|r|}
     \hline
      \textbf{Model}   & \textbf{APR EoS} & \textbf{SLy EoS} & \textbf{WFF1 EoS}
      \\  \hline
      \textbf{GR} & $M_{APR}= 1.0261\, M_{\odot}$ & $M_{SLy}= 1.033\, M_{\odot}$ & $M_{WFF1}= 1.025\, M_{\odot}$
      \\  \hline
\textbf{GR} & $R_{GR}= 11.345$km & $R_{GR}= 11.874$km
      &$R_{GR}= 10.351$km \\  \hline
      \textbf{Universal Attractors} & $M_{APR}= 1.092\, M_{\odot}$ & $M_{SLy}= 1.079\, M_{\odot}$ & $M_{WFF1}= 1.075\, M_{\odot}$
      \\  \hline
       \textbf{Universal Attractors Radii} & $R_{UNI}= 11.463$km &
$R_{UNI}= 11.827$km
      &$R_{UNI}= 10.590$km \\  \hline
    \end{tabular}
  \end{center}
\end{table}

\begin{figure}[h!]
\centering
\includegraphics[width=23pc]{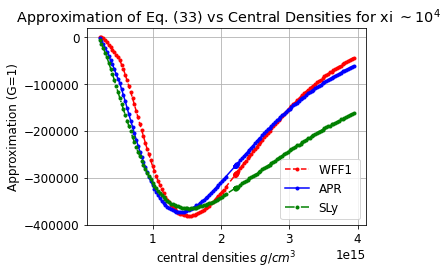}
\includegraphics[width=23pc]{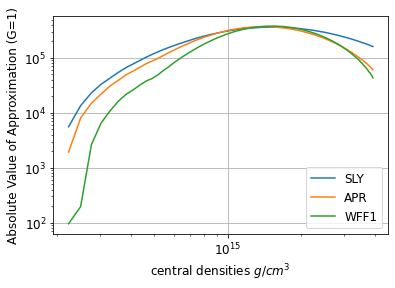}
\caption{ The ratio $\frac{12\xi^2\phi^2}{1-\xi\phi^2}$ versus the
central densities for the WFF1 EoS, the APR EoS and the SLy EoS,
and for the values of the scalar field at the surface of the NS.
As it can be seen the constraint of Eq.
(\ref{mainassumptionuniversal}) is satisfied. The bottom plot
corresponds to the absolute value of the ratio
$\frac{12\xi^2\phi^2}{1-\xi\phi^2}$ in logarithmic scale.}
\label{plot3}
\end{figure}

 In Fig. \ref{plot1} we present the $M-R$
graphs of the universal attractor models when compared to the
corresponding GR curves, for the WFF1 EoS (upper), for the APR EoS
(middle) and for the SLy EoS (bottom plot). In order to produce
Fig. \ref{plot1} we numerically solved the TOV equations and we
extracted the masses and radii corresponding to 160 central
densities, namely $M(\rho_c)$ and $R(\rho_c)$ and then we
generated the $M-R$ graph using the resulting values for the mass
and radii for each of these 160 central densities. Also in Fig.
\ref{plot2} we compare the universal attractors $M-R$ curves, for
the three distinct EoSs. A first interesting result is that the
WFF1 EoS for the universal attractor model is  compatible with the
GW170817 constraint derived in Ref. \cite{Bauswein:2017vtn}, which
indicates that $M\sim 1.6 M_{\odot}$ NSs must have radii in the
range $R=10.68^{+15}_{-0.04}$km. This is in contrast to the GR
case, where the WFF1 EoS is excluded by the GW170817 data.
Secondly, in all the studied cases, the maximum mass NS
configurations for the three distinct EoSs, satisfy the second
constraint of GW170817 derived in Ref. \cite{Bauswein:2017vtn},
which indicates that the maximum mass configurations must have
radii larger than $R=9.6^{0.14}_{-0.03}$km. Also in Table
\ref{table2} we present the radii of the static NS for all the EoS
in both GR and the universal attractors models, for which
$\frac{M}{M_{\odot}}\simeq 1$. Note that the limit $R\to 0$ is
never reached by the neutron stars, because these are GR objects
bound from gravity solely even in the context of scalar-tensor
gravity. In contrast, strange stars could reach the limit of very
small radii. Furthermore, in Table \ref{table1} we gather the data
for the maximum masses and the corresponding radii, for the GR NS
and the universal attractors NS, and for all the EoSs studied in
this paper.
\begin{table}[h!]
  \begin{center}
    \caption{\emph{\textbf{Maximum Masses and the Corresponding Radii of Static NS for the Universal Attractors  and for GR}}}
    \label{table1}
    \begin{tabular}{|r|r|r|r|}
     \hline
      \textbf{Model}   & \textbf{APR EoS} & \textbf{SLy EoS} & \textbf{WFF1 EoS}
      \\  \hline
      \textbf{GR} & $M_{max}= 2.18739372\, M_{\odot}$ & $M_{max}= 2.04785291\, M_{\odot}$ & $M_{max}= 2.12603999\, M_{\odot}$
      \\  \hline
      \textbf{Universal Attractors} & $M_{max}= 2.41712697\,M_{\odot}$ & $M_{max}= 2.27234095\,M_{\odot}$
      &$M_{max}= 2.32003695\, M_{\odot}$ \\  \hline
       \textbf{Universal Attractors Radii} & $R= 10.54678577$km &
$R= 10.56752764$km
      &$R= 9.9118728$km \\  \hline
    \end{tabular}
  \end{center}
\end{table}
Finally, we need to explicitly check whether the approximation of
Eq. (\ref{mainassumptionuniversal}), and in Fig. \ref{plot3} we
present the fraction of  and of
$\frac{12\xi^2\phi^2}{1-\xi\phi^2}$ versus the central densities,
for the three distinct EoSs, and for the values of the scalar
field at the surface of the star. As it can be seen, the
constraint of Eq. (\ref{mainassumptionuniversal}) is safely
satisfied for the values of the parameter $\xi$ we used in this
article. Also for brevity we did not include the case for the
values of the scalar field at the center of the star, in which
case the approximation of Eq. (\ref{mainassumptionuniversal}) is
satisfied.

\section*{Concluding Remarks}

In this paper we studied the phenomenology of NSs for universal
attractor non-minimally coupled scalar theories of inflation. The
universal attractors are known in cosmological contexts since
these provide a viable inflationary era and also belong to a
larger class of cosmological attractors that are similar to the
$R^2$ inflation. We investigated how the universal attractors can
be obtained in the strong coupling limit, that is for large $\xi$,
so the large coupling limit constraint must be satisfied by the
resulting values of the scalar field in the Einstein frame. After
demonstrating the essential features of the universal attractor
theories, we used the theoretical astrophysics context and we
found all the quantities that are involved in the Einstein frame
TOV equations. We solved numerically the TOV equations using a
double shooting method of a python 3 code, and we constructed the
$M-R$ graphs for all the different EoS we studied. The resulting
numerical values for the masses and radii of the NSs were the
Jordan frame ones, calculated from the resulting Einstein frame
quantities delivered by the numerical code. The results of our
analysis are interesting since we demonstrated that the WFF1 EoS
which was excluded by the GW170817 data in the context of GR, it
is not anymore excluded for the universal attractors model, since
for a NS of mass $M\sim 1.6 M_{\odot}$, the predicted radii for
the universal attractor models are larger than
$R=10.68^{+15}_{-0.04}$km. Also all the three distinct EoS for the
universal attractor models, predict higher maximum radii compared
to the GR ones, and moreover all the radii respect the constraint
of the GW170817 event which indicates that the radii corresponding
to the maximum masse must be larger than $R=9.6^{0.14}_{-0.03}$km.
A crucial issue we did not address is related to the question
whether the attractor property satisfied by the inflationary
theories, is also satisfied by the NSs. Work is in progress in
this line of research.

Finally, let us comment that the inflationary era obviously has no
effect on the neutron stars for the universal attractor potential.
Obviously the two eras are not connected, since the scalar field
value during inflation is much larger compared to the values of
the scalar field in and outside of the neutron star. The only
constraint coming from the inflationary era is on the parameter
$V_0$, the constant coupling of the scalar potential, and this is
required satisfy Eq. (\ref{tilde}).

\section*{Acknowledgments}

I am indebted to N. Stergioulas and his MSc student Vaggelis
Smyrniotis for the many hours spend on neutron star physics
discussions and for sharing his professional knowledge on
numerical integration of neutron stars in python.

\end{document}